# Multistate ferroelectric diodes with high electroresistance based on van der Waals heterostructures

*Soumya Sarkar[1*], Zirun Han[2], Maheera Abdul Ghani[1], Nives Strkalj[3], Jung Ho Kim[1], Yan Wang[1], Deep Jariwala[2], Manish Chhowalla[1*]*

[1]Department of Materials Science and Metallurgy, University of Cambridge, 27 Charles Babbage Road, Cambridge CB3 0FS, United Kingdom

[2] Department of Electrical and Systems Engineering, University of Pennsylvania, Philadelphia, Pennsylvania 19104, United States

[3]Center for Advanced Laser Techniques, Institute of Physics, 10000 Zagreb, Croatia

[*]Correspondence should be sent to ss2806@cam.ac.uk and mc209@cam.ac.uk

**ABSTRACT**

Some van der Waals (vdW) materials exhibit ferroelectricity, making them promising for novel non-volatile memories (NVMs) such as ferroelectric diodes (FeDs). $CuInP_2S_6$ (CIPS) is a well-known vdW ferroelectric that has been integrated with graphene for memory devices. Here we demonstrate FeDs with self-rectifying, hysteretic current-voltage characteristics based on vertical heterostructures of 10-nm-thick CIPS and graphene. By using vdW indium-cobalt top electrodes and graphene bottom electrodes, we achieve high electroresistance (on- and off-state resistance ratios) of $\sim 10^6$, on-state rectification ratios of 2500 for read/write voltages of 2 V/0.5 V and maximum output current densities of 100 A/cm². These metrics compare favourably with state-of-the-art FeDs. Piezoresponse force microscopy measurements show that stabilization of intermediate net polarization states in CIPS leads to stable multi-bit data retention at room temperature. The combination of two-terminal design, multi-bit memory, and low-power operation in CIPS-based FeDs is potentially interesting for compute-in-memory and neuromorphic computing applications.

**KEYWORDS:** *van der Waals heterostructure, ferroelectric diodes, $CuInP_2S_6$, multi-bit storage, electroresistance, non-volatile memory*

The rapid advancement of data-driven technologies will require the development of low-power devices that overcome the von Neumann bottleneck for efficient data processing.[1] Non-volatile memories (NVMs) based on ferroelectric materials represent a promising solution for realizing such integrated devices.[2-4] Two-terminal devices such as ferroelectric diodes (FeDs) with a metal-ferroelectric-metal vertical heterostructure and resistive readout are promising for high areal storage density and low-power operation.[3, 5, 6] Furthermore, due to their rectifying characteristics, FeDs can help alleviate the sneak-path issue without an additional selector circuit or transistor, making their crossbar arrays reliable and energy efficient.[3] The resistance states in FeDs arise from ferroelectric polarization switching induced modulation of the average barrier height (ABH). This leads to hysteretic leakage current across the ferroelectric layer – which is rectified due to the presence of a Schottky barrier at the metal-ferroelectric interface.

A robust ferroelectric diode with a high electroresistance (defined as the ratio of resistance in the on- and off- states) can provide access to multiple conductance states due to multilevel polarization switching.[7, 8] However, achieving high on- and off-state (ON/OFF) ratio requires effective modulation of the Schottky barrier height. Excellent FeD performance has been achieved with oxide and nitride ferroelectrics.[5, 6, 7, 9] van der Waals (vdW) ferroelectrics, such as $CuInP_2S_6$ (CIPS), which exhibit robust out-of-plane ferroelectricity at room temperature are also promising for FeDs.[10, 11, 12] vdW materials have a layered structure with weak interlayer bonding, and electronically and atomically sharp interfaces free of dangling bonds – facilitating the formation of heterostructures with other two-dimensional (2D) materials such as graphene.[11, 13] CIPS is a typical vdW ferroelectric in which the ferroelectric polarization is attributed to the displacement of $Cu^+$ ions within the sulphur octahedra (see schematic in **Figure 1a**).[14] It exhibits intriguing properties such as coexistence of out-of-plane ferroelectricity and ionic conductivity,[15] negative piezoelectricity,[16] and bulk photovoltaic effect.[17] Previous reports have demonstrated FEDs with thicker CIPS flakes (>30 nm).[12, 18, 19] However, it has been shown that a ferroelectric thickness of ~10 nm is the optimal regime to combine the effects of barrier height modulation and depletion width modulation to maximize the ON/OFF and rectification ratios.[2, 20]

In this study, we fabricated 8 – 10 nm thick FeDs consisting of metal/CIPS/graphene vertical heterostructures. For the top electrode (TE) we used vdW metal contacts based on an indium-cobalt (InCo) alloy, which minimise defects and Fermi level pinning at the metal-ferroelectric interface to ensure efficient carrier injection.[21] For the bottom electrode (BE), we use graphene as its Fermi level can be modulated by ferroelectric polarization switching in CIPS.[22] This asymmetric electrode device architecture allows high ON/OFF (~$10^6$) and rectification ratios (~2500) to be achieved. The large ON/OFF ratios provide robust noise immunity during operation and enable multiple conductance states (up to 5 demonstrated) that could be useful for multi-bit computing. Each of these states demonstrates robust retention for up to 100 s. We tested two states which are stable for >1000 switching cycles and can retain data for >10000 s. In addition, FeDs show high readout current densities of 100 A/cm$^2$ (maximum) and low read and write voltages of 0.5 V and <2.5 V respectively.

The CIPS flakes were mechanically exfoliated from a bulk crystal and transferred onto pre-patterned bottom electrodes on SiO$_2$ substrates using a dry transfer technique (as described in the Methods section). The thickness of the flakes was confirmed via atomic force microscopy (AFM) (see **Figure S1** in supporting information). **Figure 1b** displays the Raman spectra of CIPS on SiO$_2$ for flake thicknesses of 45 nm and 8 nm. The peak at 101 cm$^{-1}$ corresponds to the anionic (P$_2$S$_6^{4-}$) vibrations, while the peaks at 161 cm$^{-1}$, 263 cm$^{-1}$, and 374 cm$^{-1}$ are attributed to the S-P-P, S-P-S, and P-P modes, respectively.[23] The peak at 316 cm$^{-1}$, associated with the cationic (In$^{3+}$ and Cu$^+$) vibrations, is a signature of the ferroelectric phase.[23, 24] All these modes are present in the 8-nm-thick CIPS flake, confirming the ferroelectric crystal structure.

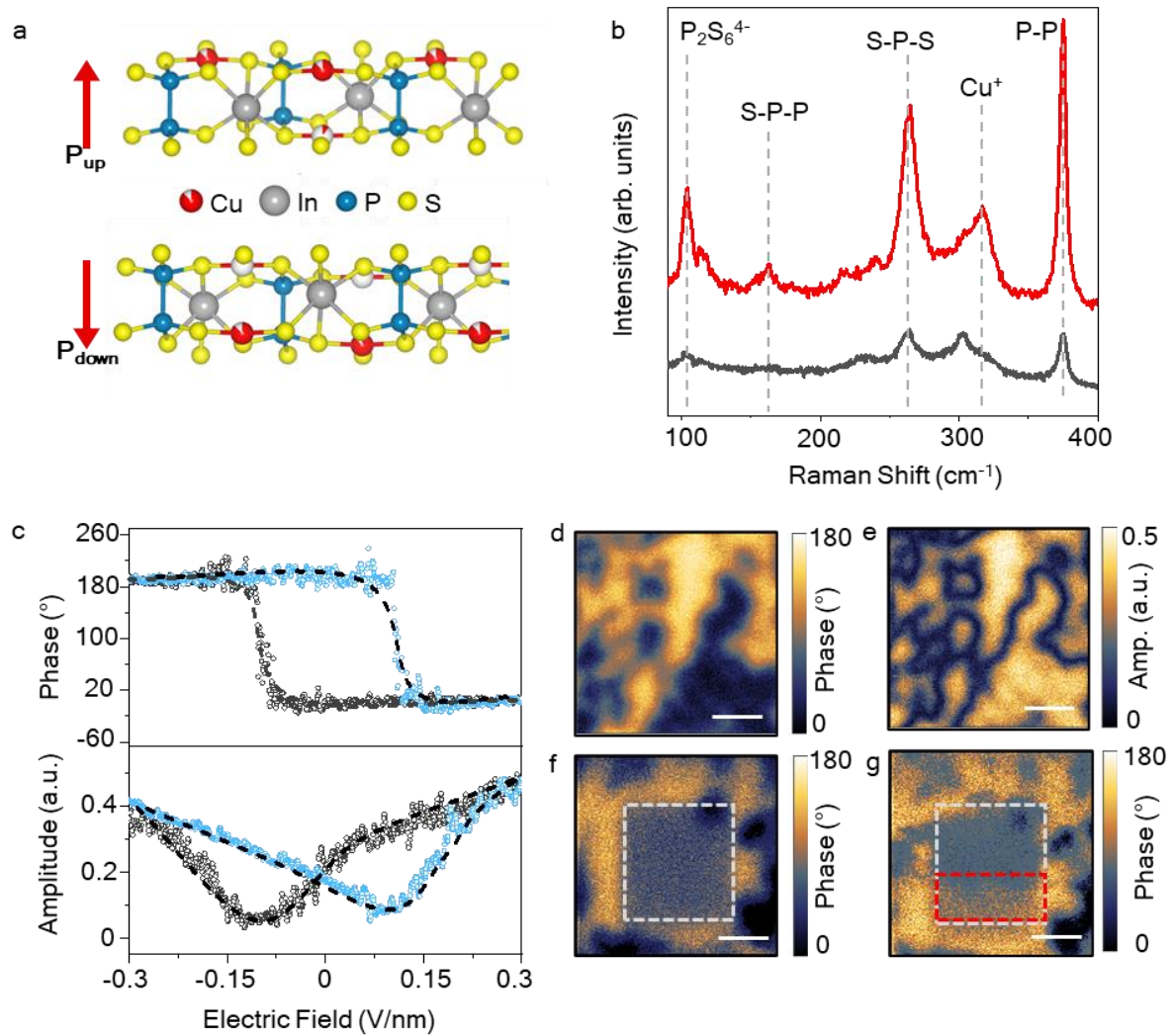

**Figure 1 Ferroelectricity in ultrathin CIPS** (a) Schematic of crystal structure of $CuInP_2S_6$ (CIPS) for the two opposite polarization configurations originating from the displacement of the $Cu^+$ ion (red) (b) Raman spectra of mechanically exfoliated CIPS flake on $SiO_2$/Si substrate for flake thickness of 45 nm (red) and 8 nm (grey). The characteristic ferroelectric Raman mode ($Cu^+$) is detected at 8 nm flake thickness. (c) Out-of-plane PFM phase and amplitude loops show characteristic ferroelectric hysteresis measured on a 20-nm-thick CIPS/Au test structure. Out-of-plane PFM (d) phase and (e) amplitude maps of the same as-exfoliated CIPS flake shows presence of polar domains. (f) The ferroelectric polarization switching is visualized by PFM images after writing a square region with -2.5 V (white box) and (g) +1 V, which causes partial polarization reversal (red box). The scale bar represents 500 nm.

The ferroelectric properties of CIPS were characterized using piezoresponse force microscopy (PFM). **Figure 1c** shows the out-of-plane PFM phase and amplitude measurements on a 20-nm-thick CIPS/Au/SiO$_2$ test device. The clear 180° switching of the PFM phase signal and the well-defined butterfly loops in the PFM amplitude signal indicate robust ferroelectric polarization switching.[12] Furthermore, the PFM phase and amplitude images of the pristine CIPS sample have been mapped (**Figure 1d, e**), confirming the presence of multiple ferroelectric domains with a lateral size of ~200 nm. The PFM amplitude is constant within the domains and reduced at the domain walls, characteristic of ferroelectric films. The PFM phase reveals domains characterized by two 180°-spaced contrast levels, corresponding to the two opposite polarization directions perpendicular to the film surface. **Figure 1f** shows the PFM phase images of the same CIPS flake after writing a square pattern (white dashed box) with negative DC bias of -2.5 V. By applying a positive DC bias of +1 V, we obtain partial reversal of the polarization (phase contrast in the red dashed box in **Figure 1g**), suggesting stable intermediate net polarization states in CIPS. Such gradual manipulation of ferroelectricity has previously been observed in oxide ferroelectric thin fims.[8]

We have investigated the electrical properties of metal/CIPS/metal vertical heterostructures, as shown in **Figure 2**. The current-voltage characteristics were analysed for CIPS flakes with thicknesses ranging from 8 nm to 10 nm, sandwiched between electrodes in a vertical crossbar geometry. The blue plot corresponds to the forward sweep direction from negative to positive bias and the red plot represents the reverse sweep direction. The BE was electrically grounded for all measurements.

In the first configuration, the BE is a lithographically patterned 20-nm-thick Au strip and the TE is metallic vdW contact based on an In/Co alloy. The In/Co alloy contact was capped with 5 nm thick Au to prevent oxidation (details of the device fabrication process is discussed in the methods section). For the Au/CIPS/InCo vertical heterostructure (**Figure 2a**, with schematic shown as inset), hysteretic and non-linear current-voltage characteristics were observed (the linear current-voltage characteristics are presented in **Figure S2** of supporting information). The observed non-linearity in resistive switching, with an ON state rectification ratio of ~20 at 0.5 V read voltage, can be attributed to the polarization related changes in ABH due to asymmetric metal contacts (Au vs. In/Co).

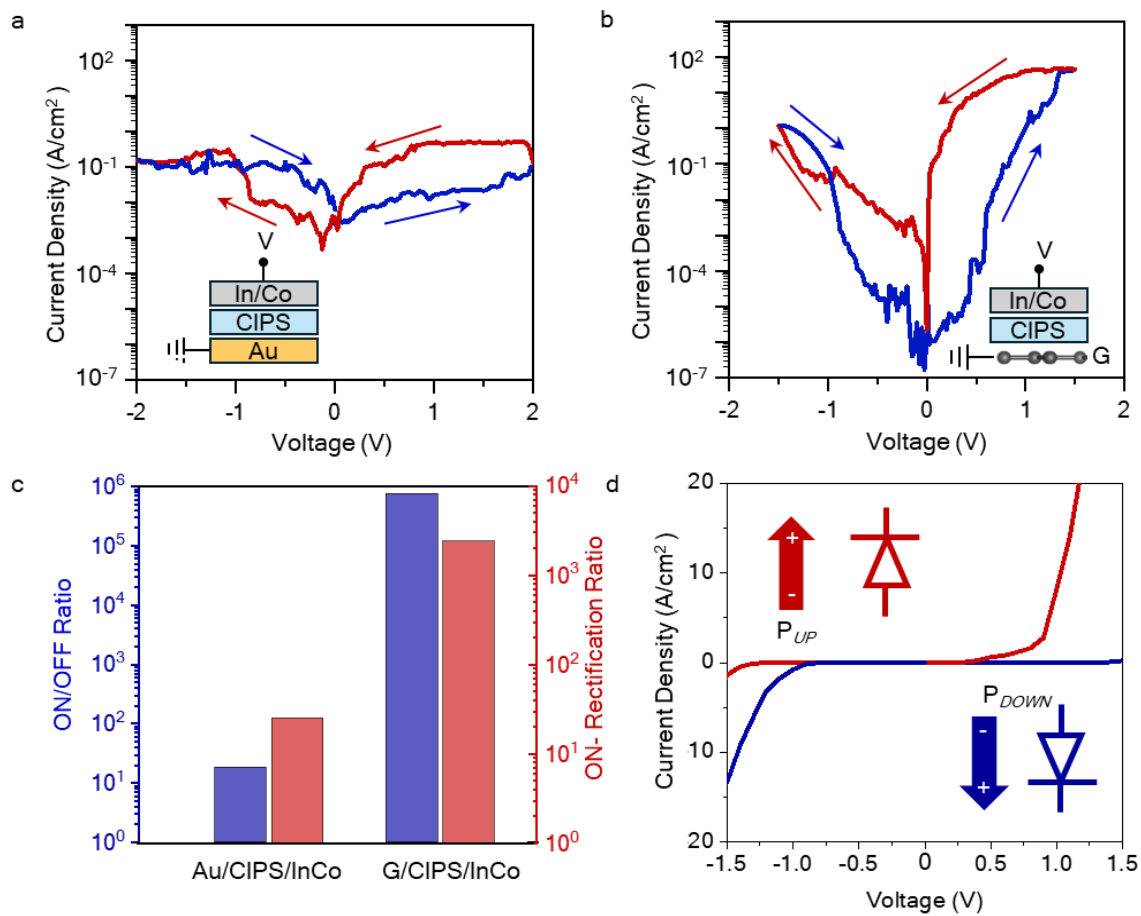

**Figure 2 Electrical characteristics of CIPS based FeDs** Current density vs voltage characteristics measured for two terminal CIPS-based crossbar devices with asymmetric electrodes – (a) Au and In/Co, and (b) In/Co and graphene electrodes. The cell size varies from 0.7 - 2 μm$^2$. The schematics of the two devices are shown in the inset. The blue and red arrows represent the forward and reverse voltage sweep directions. (c) ON/OFF ratio (left axis) and rectification ratio (right axis) for devices with three different electrodes at 0.5 V read voltage. (d) Nonlinear diode-like current-voltage characteristics of a G/CIPS/InCo device with a cell size of 1 μm$^2$.

To further investigate the influence of asymmetric Schottky barriers at the interface, we used graphene strips as the bottom electrode (BE) and In/Co as the top electrode (TE). The devices were fabricated by mechanical transfer of CIPS flakes on 1-2-layer-thick graphene on SiO$_2$ (described in methods). The G/CIPS heterostructures were characterized by Raman spectroscopy and Raman mapping to identify any heterogeneities in the transfer process (see **Figure S3** of supporting information). For the G/CIPS/InCo vertical heterostructure devices (**Figure 2b**, with schematic and OM image in inset), we observed a similar counterclockwise hysteretic current-voltage characteristic, with a giant enhancement in ON/OFF and rectification ratios (the linear current-voltage characteristics are presented in **Figure S2** of supporting information). We have presented the ON/OFF ratios and the ON state rectification ratios of the three devices at a low read voltage of 0.5 V in **Figure 2c**. The variation of ON/OFF ratios of the three devices as a function of applied voltage is shown in **Figure S4** of supporting information. The device with graphene BE has an ON/OFF ratio of approximately $10^6$ at 0.5 V and a rectification ratio of around 2500 (defined as the ratio of the current in the ON state at +0.5 V and -0.5 V). These values are several orders of magnitude higher than devices with metallic BEs, and at least an order of magnitude greater than those reported for FeDs based on traditional perovskite oxides or $HfO_2$.[20, 25] Furthermore, the current density in the OFF state is reduced in our G/CIPS/InCo FeD devices in comparison with Au/CIPS/InCo FeD devices suggesting transport is limited at the G/CIPS interface. A high readout current density of 100 $A/cm^2$ is measured in the ON state, which is advantageous for high-frequency readout when displacement currents, which scale with frequency, typically become dominant.

The current-voltage curve (in linear scale) of the FeD shown in **Figure 2d** represents the characteristics of a diode. After sweeping to positive voltages, the current changes from low to high, and the diode polarity shifts from an OFF-forward diode to an ON-forward diode (red line). When the voltage sweep direction is reversed towards negative voltages, the diode polarity reverses from an ON-reverse diode to an OFF-reverse diode (blue line), as shown in the inset.

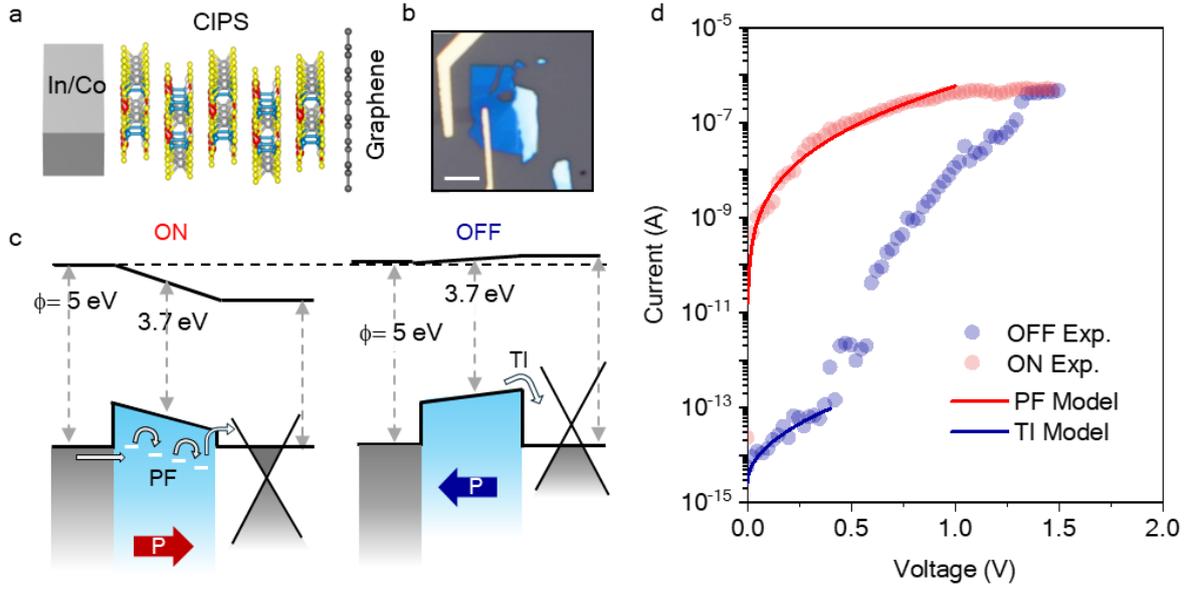

**Figure 3 Transport mechanism in CIPS based FeDs** (a) Schematic structure and (b) optical image of the G/CIPS/InCo FeD. Scale bar is 5 µm. (c) Energy band diagrams and active transport mechanisms of the FeD for ON (left) and OFF state (right). The diagram shows the ferroelectric polarization switching induced ABH modulation and dominant transport mechanisms. (d) Current-voltage characteristics of the FeD fitted with the Poole-Frenkel (PF) and thermionic emission (TI) models.

**Figure 3a** shows the schematic of the 10-nm-thick CIPS based FeDs, where Schottky contacts are formed at G/CIPS and CIPS/InCo interface (optical image of the device is shown in **Figure 3b**). The observed non-linearity, in our current-voltage characteristics is due to the modulation of the ABH and/or depletion layer width in forward bias and reverse bias regimes. The modulation of ABH is due to the opposite polarization charges at the ferroelectric-electrode interface for the two polarization states which can lead to different conduction mechanisms through the device.[26] As shown in **Figure 3c**, the high ON/OFF ratio is related to graphene electrode. The low quantum capacitance of graphene near the Dirac point allows for effective modulation of its Fermi level as a result of ferroelectric polarization switching in CIPS.[22] In the ON state, the polarization of CIPS points towards the graphene BE and modifies graphene from an intrinsic to an n-type semiconductor, raising the Fermi level above the Dirac point (left side of **Figure 3c**). This increase in the Fermi level reduces ABH, making it easier for electrons to propagate through CIPS, resulting in higher current. The thickness ~10 nm of the CIPS layer means that direct tunnelling is limited, and conduction across the ferroelectric can occur through defect-assisted tunnelling or thermionic injection of carriers above the barrier.[27] Based on the high output current in the ON state, the dominant conduction can be accounted for by Poole-Frenkel (PF) tunnelling – as confirmed by the fit using PF model (see red curve in **Figure 3d**, and description of fitting model in the methods section).[7, 9] The PF conduction mode is likely enabled by the presence of point defects in the CIPS crystal. We additionally observe that the ON current saturates at larger biases, and this behaviour can be attributed to high-field saturation in the PF mechanism.[28]

In the OFF state, the polarization in CIPS points in the opposite direction, shifting the Fermi level of graphene below the Dirac point to p-type (right side of **Figure 3c**). This increases the ABH and leads to low OFF state current (~$10^{-14}$ A). From **Figure 3d**, we attribute the interface-limited thermionic emission (TI) over the G/CIPS barrier as the current limiting factor in the OFF-state current. The transport mechanism in the negative voltage regime of the FeD and the Au/CIPS/InCo FeD is described in **Section 5** of the supporting information.

The current–voltage characteristics of the FeD (shown in **Figure 4a**) indicate that the transition between the OFF state and the ON state has a low switching slope (~0.25 V/dec), suggesting a gradual polarization reversal in the CIPS. As shown in the PFM images in **Figure 1d, e** above, the ferroelectric domain size in pristine CIPS is ~200 nm, which suggests presence of a multiple polar domains in a memory cell of area 1-2 $\mu m^2$. Therefore, on applying sub-coercive voltages, intermediate net polarization states can be stabilised due to different domain populations as shown by PFM in **Figure 1g**. Such intermediate polarization states can reflect as multiple resistance states to store information.[12]

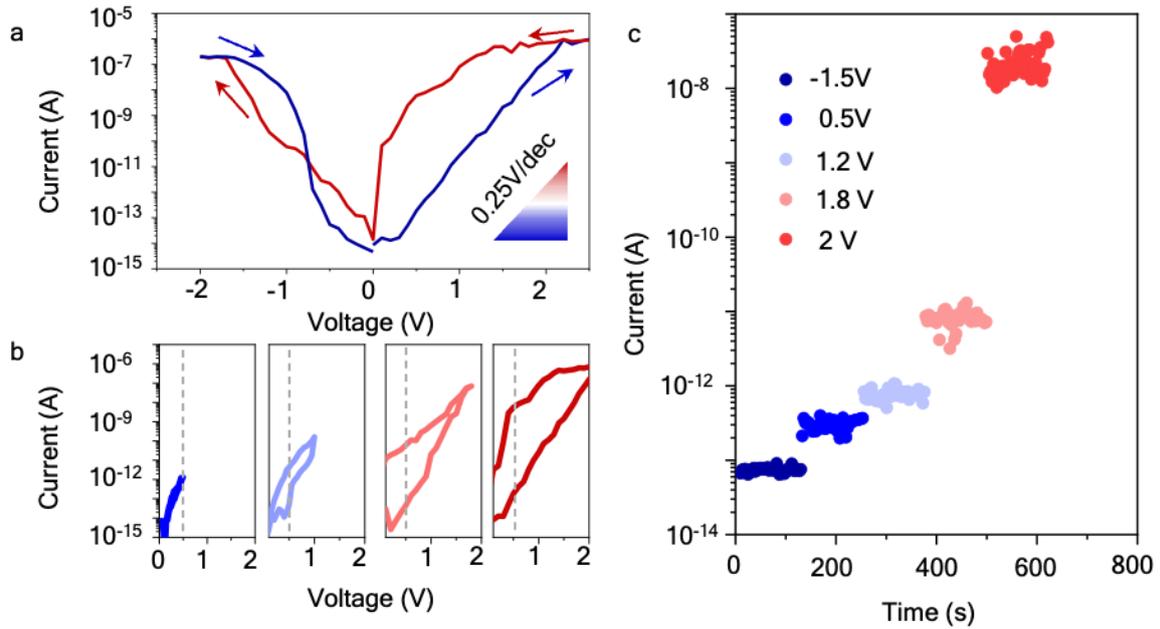

**Figure 4 Multiple retention states in CIPS based FeDs** (a) Current-voltage characteristics of the FeD shows a gradual polarization switching from the OFF state (blue) to ON state (red). (b) Current-voltage characteristics measured across a voltage sweep range of 0-0.5 V, 1 V, 1.8 V, 2.2 V show emergence of hysteresis due to partial polarization switching. Prior to these sweeps, the FeD was reset to the OFF state by applying -1.5 V. (c) Multistate retention of the FeD up to 100s for 5 states measured at 0.5 V read voltage (grey dashed line in (b)).

To stabilise these intermediate resistance states, we gradually increase the voltage scan range from 0 – 0.5 V, 0 – 1 V up to 2.5 V. As shown in **Figure 4b**, the memory window related to ferroelectric hysteresis increases as the scan range increases, allowing us to read multiple intermediate non-volatile resistance states at a read voltage of 0.5 V (grey dashed line). In **Figure 4c**, we demonstrate multistate retention for over 100 s for five stable intermediate resistance states, though the state separation would clearly allow for a higher number of distinguishable states. These multiple resistance states are related to the different net polarization states in CIPS.

**Figure 5a** illustrates the data retention of two of these resistance states. Prior to measuring the retention, the device was first set to the OFF/ON state by applying a write/erase voltage of -2V/+2.5V respectively. Subsequently, the data was read with a bias of 0.4 V. The ON/OFF ratio of the device remained above $10^5$ after more than 10000 s, demonstrating potential for achieving long data retention times in both the ON and OFF states. In **Figure 5b**, we show the endurance measurement on this FeD over 1500 cycles. Within each cycle, the FeD is set to the ON state by a +2.5 V pulse (10 ms) and read at a bias of 0.5 V, then set to the OFF state using a -2 V pulse (10 ms) and read at a bias of 0.4 V. The ON/OFF ratio remains above 3000 after 1000 write/erase cycles.

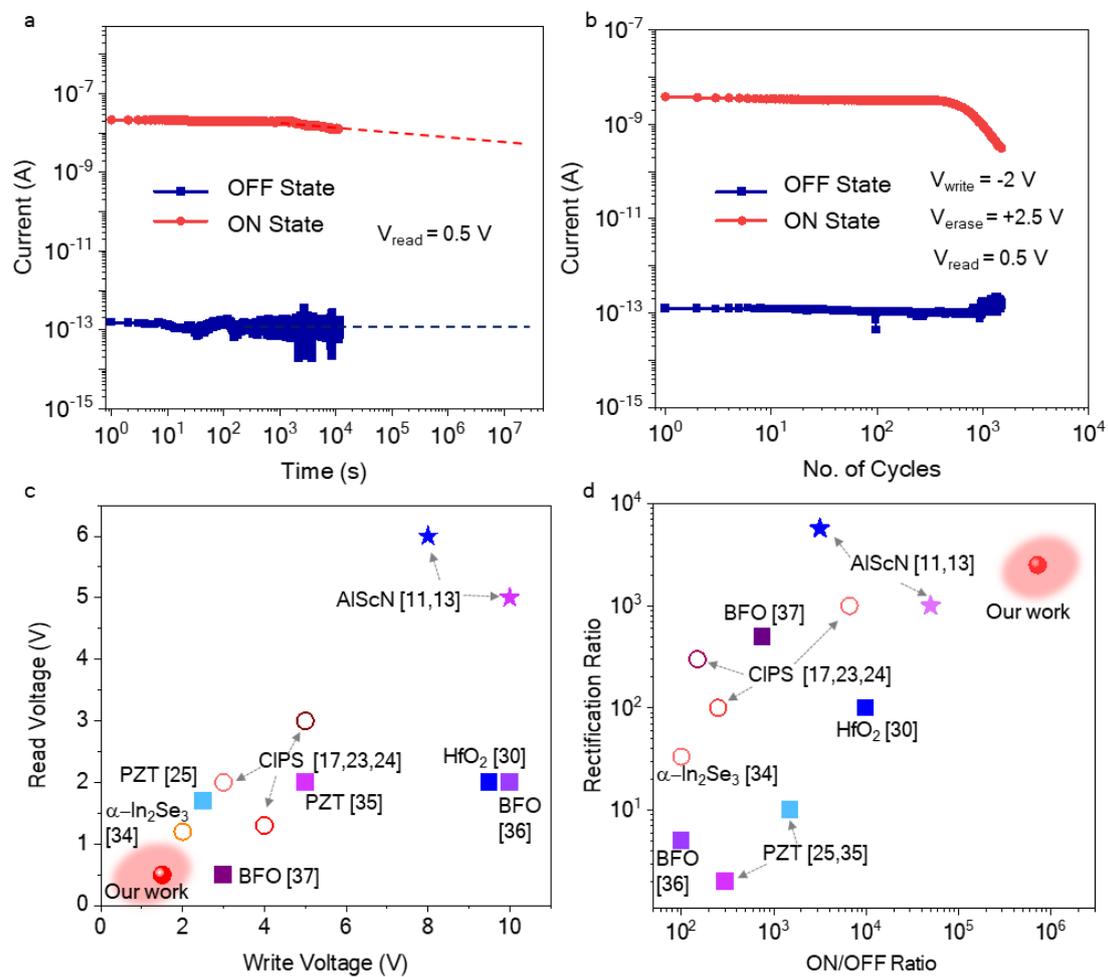

**Figure 5 Memory performance of CIPS based FeDs** (a) Two state data retention measurements of 10 nm-thick CIPS based FeD for 12000 s. (b) The endurance characteristic of the device. Experimental measurements were obtained for 1600 switching cycles. Comparison of the (c) read voltage versus write voltage and (d) rectification ratio versus ON-OFF ratio of CIPS based FeDs with previous reported FeDs in literature. The measurements were performed at room temperature.

A comparison of the read and write voltages, and ON/OFF ratio and rectification ratio of our CIPS based FeDs with previously reported FeDs is presented in **Figure 5c,d**.[5, 9, 12, 15, 19, 25, 29] The sample thickness for the above reports and comparison with 3 CIPS based FeDs is presented in Table S1 of supporting information. The electrical characteristics of our devices compares favourably with other reports. Furthermore, the memory performance is improved over previous FeDs based on vdW ferroelectrics. The electrical properties of all our devices as compared to FeDs reported in literature are summarized in **Section S6** of supporting information.

In summary, we have demonstrated FeDs based on vertical heterostructures of layered ferroelectric CIPS with graphene and In/Co vdW metal contacts. The high electroresistance and rectification ratios observed in our devices are attributed to the efficient polarization-modulated transport at G/CIPS contact interface. Our results indicate that vdW ferroelectric diodes are promising for the design of high-density, selector-free embedded NVMs capable of efficient data processing. We have demonstrated stable data retention over multiple states, which could be beneficial for the development of neuromorphic hardware. In future, lateral scaling of devices and large area, thickness controlled scalable growth strategies of the CIPS material will be necessary for advancing this promising device concept into a memory technology.

**ASSOCIATED CONTENT**

**Supporting Information**

The Supporting Information is available free of charge. It contains the following information: Experimental details, Atomic force microscopy (AFM) images of G/CIPS device, linear current-voltage characteristics of FeDs, Raman spectroscopy of G/CIPS vdW heterostructures, voltage dependence of ON/OFF ratios, conduction mechanism of FeDs, and comparison with previously reported FeDs

AUTHOR INFORMATION

**Corresponding Authors**

**Soumya Sarkar -** *Department of Materials Science and Metallurgy, University of Cambridge, 27 Charles Babbage Road, Cambridge CB3 0FS, United Kingdom*

Email: ss2806@cam.ac.uk

**Manish Chhowalla** - *Department of Materials Science and Metallurgy, University of Cambridge, 27 Charles Babbage Road, Cambridge CB3 0FS, United Kingdom*

Email: mc209@cam.ac.uk

**Author Contributions**

S.S. prepared and characterized samples, fabricated, measured devices, and interpreted results with help from M.A.G., N.S., J.H.K., and Y.W. Z.H. modelled the device transport mechanisms under the supervision of D.J. M.C. supervised the project. S.S., D.J., and M.C. wrote the manuscript. All authors read the paper and agreed on its content.

**Notes**

The authors declare no competing financial interest.

ACKNOWLEDGEMENTS

We acknowledge funding from European Research Council (ERC) Advanced Grant under the European Union's Horizon 2020 research and innovation programme (grant agreement GA 101019828-2D- LOTTO]), EPSRC (EP/ T026200/1, EP/T001038/1). D.J. and Z.H. acknowledge primary support from the Air Force Office of Scientific Research (AFOSR) GHz-THz program FA9550-23-1-0391. Z.H. also acknowledges funding support from the Vagelos Integrated Program in Energy Research (VIPER) at Penn.